# Engineering Morphologies of Metal-Based Colloidal Assemblies via Colloid Jamming at Liquid-Liquid Interfaces

Jiyuan Yao,[1,3] Shuting Xie,[1] Shijian Huang,[2] Weilong Xu,[1] Jiaqin Li,[1] Zhenping Liu,[1] Mingliang Jin,[1] Loes I. Segerink,[3] Lingling Shui,[1,2]* and Sergii Pud[3]

[1]*International Joint Laboratory of Optofluidic Technology and System (LOTS), National Center for International Research on Green Optoelectronics, Guangdong Provincial Key Laboratory of Nanophotonic Functional Materials and Devices, South China Academy of Advanced Optoelectronics, South China Normal University, Guangzhou, 510006 (China)*

[2]*School of Optoelectronic Science and Engineering, South China Normal University, Guangzhou, 510006 (China)*

[3]*BIOS Lab-on-a-chip group, EEMCS Faculty, MESA+ institute, University of Twente, Enschede, 7500 AE (The Netherlands)*

*Email address: shuill@m.scnu.edu.cn

# Abstract

Self-assemblies, structured *via* nanoparticles, show promise as materials for advanced applications, like photonic devices, electrochemical energy storage units and catalysis support. Despite observing diverse morphologies, a comprehensive understanding of the formation mechanism remains elusive. In this work, we show that the coordination interaction between metal-based sulfide nanoparticles (MS NPs) and the fluorosurfactants at the droplet interface influences the morphology during the evaporation-induced self-assembly facilitated by droplet microfluidics. Further investigation into fluorosurfactants with various chemical groups and MS NPs reveals that the strength of coordination interactions significantly influences assembly morphology. The interfacial interactions can be eliminated through coating a $SiO_2$ layer on the metal-based colloid (M@$SiO_2$ NPs). In addition, we demonstrate that the morphologies of the self-assemblies can be engineered *via* the coordination interactions between the MS NPs and fluorosurfactants, and by varying the concentrations of MS NPs. Utilizing these interfacial interactions, assemblies with core-shell and homogeneous distribution of binary nanoparticles were constructed. Our findings offer novel insights into the interfacial jamming of nanoparticles at the droplet interface through evaporation-induced self-assembly, and into the design of metal-based colloidal assemblies with diverse morphologies, crucial for developing novel functional assemblies for catalysis, plasmonic, and porous materials in a controlled manner.

KEYWORDS: Metal-based colloids, Self-Assembly, Interfacial jamming, Emulsions, Fluorosurfactants

# Introduction

Colloidal self-assemblies are powerful and versatile tools for engineering new (micro) materials with applications across various domains of science. They offer tunability of the components, control of size, and possibility of building hierarchical structures[1-4]. Colloidal self-assemblies find applications as catalytic supports[5-7], controllable superwetting modulation[8,9], sensing as substrates[10,11], and optoelectronic devices[12,13], metamaterials[14,15], drug delivery[16,17]. Colloidal self-assemblies are created by confining nanoparticles (NPs) within a specific template and further consolidating them into supraparticles (or patterns). Droplet microfluidics is an essential tool for producing confinement units enabling size-controlled and highly uniform assemblies in microdroplets. Compared to the traditional confinement units such as sessile droplets and microwells, microdroplets offer minimal contact with the substrate, easy sample recovery, and highly controllable self-assembly process[18,19]. The second step of self-assembly requires consolidation of the confined colloids, which can be done through polymerization induced by UV light, ion irradiation, thermal treatment or solidification *via* evaporation. The latter has garnered substantial attention for assemblies, due to its simplicity, cost-effectiveness and robustness[20]. During the solidification process, microdroplets serving as the dispersed phase are segregated by an oil phase enriched with surfactants.

In conventional colloidal assemblies structured by evaporation, such as those involving silica nanoparticles ($SiO_2$ NPs)[15,21] or polystyrene nanoparticles (PS NPs)[22,23], the self-assembly typically unfolds due to the interparticle interactions within the template[24,25]. These interactions are well studied and are shown to be primarily governed by the balance of van der Waals force, gravitational forces, capillary forces, and electrostatic forces, which determines the final packing of the colloids based on their interaction in suspension[26-28]. For example, Butt et al.[29] revealed that the external shape and the internal structure of the self-assembly could be controlled through tuning the electrostatic interaction between primary colloids. Nicolas et al.[30] demonstrated that the evaporation rate significantly influences the shape and crystalline structure of assemblies, varying from buckled to spherical forms, and evolving from local order with partial icosahedral to fully icosahedral crystallinity. Zhang et al.[31] reported the shapes of self-assemblies are determined by the size of particles. In addition, the interparticle interactions that govern self-assembly have also been studied in binary colloidal assemblies, involving interactions between the different types of colloids[24, 32-34]. However, understanding of the external interactions, such as those between the colloids and oil/surfactants, are still in the early stage. For example, the atypical assembly geometry, specifically concave colloidosomes, has been observed in the self-assembly of zeolitic imidazolate framework-8 (ZIF-8) within microdroplets[6, 35]. This phenomenon has been roughly explained by trapping the ZIF-8 at the microdroplet interface due to sufficiently slow solvent removal, particularly at low concentrations of ZIF-8 in the suspension. Recently, electrostatic interaction between interparticle and fluorosurfactants at the droplet interface has been revealed to block the supraparticle with different morphologies[36]. In addition, in our previous work, we demonstrated that copper-based supraparticles ($Cu_xO$ NP-SPs) and copper-based colloidosomes ($Cu_xO$ NP-CSs) could be structured *via* modulating the interactions between the $Cu_xO$ NPs and fluorosurfactants at the microdroplet interface, rather than relying solely on the physical interactions between the particles[37]. This kind of interactions between the $Cu_xO$ NPs and fluorosurfactants, results in trapping of copper-based NPs at the microdroplet interface opposed to the volume packaging. We find that the impact of the surface interactions between the colloids and the surfactant manifest itself strongly in shaping metal-based colloidal self-assemblies, which exhibit unique properties, such as enhanced porosity, high catalytic activity, and multifunctionality[20, 27, 34, 38, 39]. Understanding these interactions and studying their impact on the morphology of the resulting assemblies opens new opportunities for engineering the shape and structure of colloidal self-assemblies by tweaking the properties of the surfactant without altering the properties of the trapped colloids.

In this work, we demonstrated that the morphology of metal-based assemblies can be controlled by tuning the interfacial interactions between metal-based colloids and fluorosurfactants. Additionally, we show that the concentration of metal-based colloids in the suspension also significantly impacts the final assembly properties. Specifically, we structured clusters containing cadmium sulfide nanoparticles (CdS NPs) with colloidosomes (CdS NP-CSs) and supraparticles (CdS NP-SPs) morphologies. These morphologies were achieved with jamming or no jamming the CdS NPs at the droplet interface using fluorosurfactants PFPE(H)-Tris or PFPE(H)$_2$-ED$_{900}$. To mimic ZIF-8, we used zinc sulfide nanoparticles (ZnS NPs), which share a similar elemental composition to ZIF-8, but exhibit distinct self-assembly behavior due to their stronger interaction with fluorosurfactants compared to CdS NPs. These coordination interactions were eliminated by coating the metal-based sulfide (MS) NPs with a silica layer, forming MS@SiO$_2$ NPs (CdS@SiO$_2$ NPs and ZnS@SiO$_2$ NPs). In addition, we demonstrated the interfacial trapping of MS NPs at the drop interface in a pendant drop after solvent extraction, and characterized the interfacial interaction *via* X-ray photoelectron spectroscopy (XPS). This approach allowed us to guide the morphology of the assemblies through coordination interaction and concentration adjustment during the evaporation-induced self-assembly. Finally, we also demonstrated the role of surfactant-colloid interactions by forming and studying binary colloidal self-assemblies composed of colloids with differing affinities for the fluorosurfactants. Our findings provide fundamental understanding of the final self-assembly morphologies controlled by interactions between surfactants and metal-based colloids. We expect these results to yield additional control and flexibility for engineering materials based on colloidal assemblies, but also highlights the role of fluorinated oil combined with fluorosurfactants as the continuous phase, aiding in this assembly process.

## Results and discussion

**Tunable Morphologies of MS Self-assemblies' Construction**

A microfluidic device featuring flow-focusing geometry (**Figure 1 a-i**) was used to generate the microdroplets containing MS NPs. Different MS NPs, including CdS NPs, CdS@SiO$_2$ NPs, ZnS NPs and ZnS@SiO$_2$ NPs (size distribution and morphologies of nanoparticles are shown in **Figure S1**), were generated in fluorinated oil with different fluorosurfactants (PFPE(H)$_2$-ED$_{900}$ or PFPE(H)-Tris. **Figure S2** shows the fluorosurfactant synthesis protocol (**Figure S2 a**) and the FT-IR spectra of fluorosurfactants (**Figure S2b**). Fluorinated oil combined with a fluorosurfactant serves as an ideal continuous phase for droplet-confined colloidal self-assembly[6, 12].

As shown in **Figure 1 a-ii**, microdroplets containing 1.0 wt% CdS NPs dispersed in a water and ethanol mixture (1v/1v) were emulsified in fluorinated oil with a specific fluorosurfactant using a PDMS generator, driven by a syringe pump with a controlled flow rate. Subsequently, self-assemblies were formed *via* evaporation-induced self-assembly in an isothermal oven at 33 ºC (**Figure 1 a-iii**). We tracked the evolution of droplet shapes changes by optical microscopy. Intriguingly, distinct droplet shapes and assembly morphologies were observed after solidification under identical conditions, depending on the fluorosurfactants used to stabilize the microdroplets. The emergence of different shapes and morphologies is attributed to the interactions between the fluorosurfactants and CdS NPs at the droplet interface. In brief, during the evaporation-induced self-assembly process under the stabilization with the fluorosurfactants PFPE(H)-Tris (**Figure 1 b**), the CdS NPs were initially homogeneously dispersed within the microdroplet (**Figure 1 b-i**). As evaporation proceeded, the CdS NPs became jammed at the droplet interface due to PFPE(H)-Tris, leading to the formation of a concave microcapsule composed of CdS NPs, referred to as CdS NP-colloidosome (CdS NP-CS), was formed (**Figure 1 b-ii**). Conversely, with PFPE(H)$_2$-ED$_{900}$ as the stabilizer (**Figure 1 c**), the CdS NPs were initially homogeneously dispersed in the microdroplet as well (**Figure 1 c-i**). As evaporation continued, capillary forces from droplet shrinkage drove the CdS NPs inward, resulting in the formation of a supraparticle composed of CdS NPs, termed CdS NP-supraparticle (CdS NP-SP), was constructed (**Figure 1 c-ii**). These results demonstrate that CdS NPs were irreversibly absorbed at the droplet interface by fluorosurfactant PFPE(H)-Tris, leading to their interfacial trapping and the formation of CdS NP-CSs during self-assembly. In contrast, spherical and condensed suprapartices were formed using PFPE(H)$_2$-ED$_{900}$, due to the reversible adsorption of the CdS NPs at the droplet interface[26].

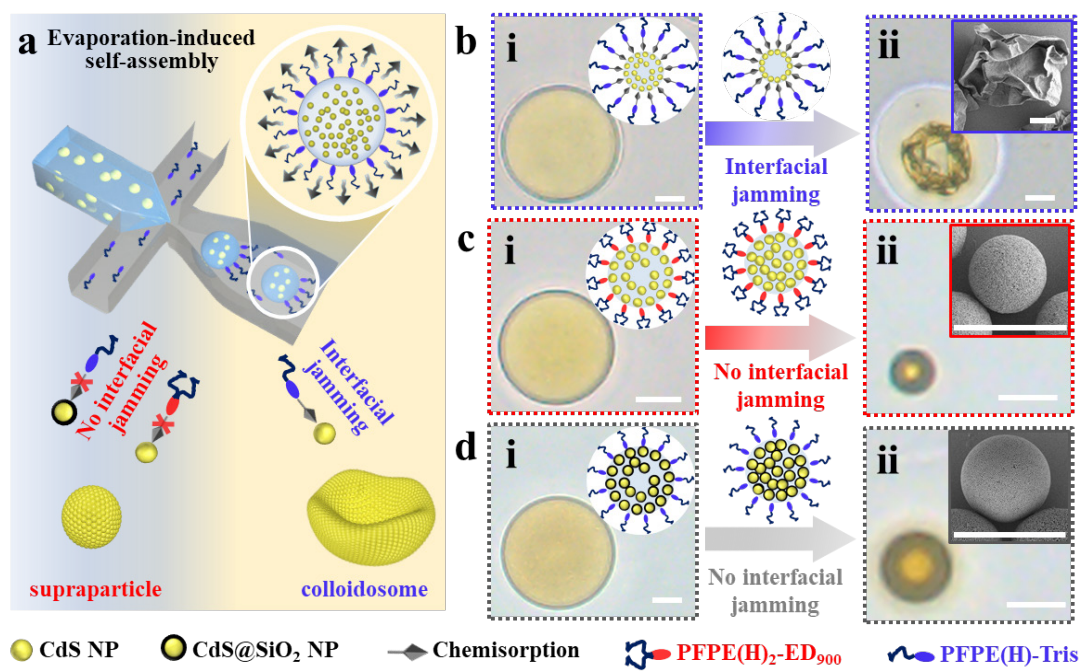

*Figure 1. Morphologies of CdS assemblies under droplet-confined self-assembly with different fluorosurfactants. a) Schematic illustration showing the effect of interfacial jamming/no jamming on CdS NP-assemblies' morphologies. b) Assembly behavior of CdS NPs in microdroplet under the interfacial jamming caused by fluorosurfactant PFPE(H)-Tris at concentration of 5.0 mM, showing optical microscopy image of emulsion droplet formation containing CdS NPs (i) and their assembly under interfacial jamming, along with a scanning electron microscopic image of CdS NP-colloidosome (ii). c) Assembly behavior of CdS NPs in microdroplet without the interfacial jamming caused by fluorosurfactant PFPE(H)$_2$-ED$_{900}$ at concentration of 5.0 mM, including optical microscopy image of emulsion droplet formation containing CdS NPs (i) and their assembly without the interfacial jamming with an SEM image of CdS NP-suprapartice (ii). d) Assembly behavior of CdS@SiO$_2$ NPs in microdroplet without the interfacial jamming caused by fluorosurfactant PFPE(H)-Tris at concentration of 5 mM, showing optical microscopy image of emulsion droplet formation containing CdS@SiO$_2$ NPs (i) and their assembly without interfacial jamming, with a scanning electron microscopic*

*image of CdS@SiO$_2$ NP-supraparticle (ii). Scale bar: 10 μm for all optical and SEM images in (b-d).*

The hydroxyl group from fluorosurfactants could interact with metal ions[40, 41], which might cause the CdS NPs to jam at the interface by PFPE(H)-Tris. To demonstrate that the metal (Cd$^{2+}$) is critical for achieving the interfacial jamming, we self-assembled the CdS@SiO$_2$ NPs — with a core of CdS NPs and shell of SiO$_2$ — in the microdroplet with fluorosurfactant PFPE(H)-Tris (**Figure 1 d**). Due to the silica layer of CdS@SiO$_2$ NPs, we expect no interaction between the metal and the surfactant. Like the CdS NPs, the CdS@SiO$_2$ NPs homogeneously dispersed in the microdroplets as well (**Figure 1 d-i**). During the evaporation-induced self-assembly, the CdS@SiO$_2$ NPs were moved into the microdroplet by capillary forces from the microdroplet shrinkage rather than jammed at the droplet interface; finally, a supraparticle consisting of CdS@SiO$_2$ NPs, named as CdS@SiO$_2$ NP-supraparticle (CdS@SiO$_2$ NP-SP), was constructed (**Figure 1 d-ii**). These results indeed confirm that interaction between metal and surfactant is needed for colloidosome formation. The time evolution of CdS NPs, CdS@SiO$_2$ NPs self-assembly with different fluorosurfactants is shown in **Video S1-S3**.

To further explore the interfacial jamming with other MS NPs, we conducted the evaporation-induced self-assembly using ZnS NPs, considering that zinc is a conventional element of metal-organic framework materials[6, 35]. The results of a series of ZnS NP-assemblies were presented in **Figure S3**, where similar patterns of interfacial jamming were observed. However, ZnS NPs exhibited stronger interfacial interaction than CdS NPs, as evidenced by the formation of ZnS NP-CS formed under stabilization by PFPE(H)$_2$-ED$_{900}$. The interfacial jamming was eliminated by coating the ZnS NPs with a SiO$_2$ layer as well.

The morphology of colloidosomes with a concave shape is rarely discussed in evaporation-induced self-assembly. This is ascribed to the interfacial jamming of nanoparticles caused by their interactions with the fluorosurfactants PFPE(H)-Tris.

However, elucidating the influence of fluorosurfactants on the evaporation-induced process is crucial for understanding colloidal self-assembly, as the resulting morphology is strongly affected by surfactant interactions.

**Mechanism of the colloid-surfactant interaction for self-assembly formation**

The pendant drop technique is commonly used to study interfacial jamming by capturing wrinkle formation at the droplet interface using a high-speed camera[42-46]. To further investigate the interfacial jamming, we tracked the evolution of drop shapes using the pendant drop technique. The droplets contained a dispersion of either CdS NPs or CdS@SiO$_2$ NPs at a concentration of 1.0 wt% in a homogeneous mixture of water and ethanol (1:1 volume ratio), and were floated in fluorinated oil containing fluorosurfactants at a concentration of 5.0 mM. As shown in **Figure 2 a**, the interfacial wrinkle was observed in the droplet containing 1.0 wt% CdS NPs after the solvent extraction, when immersed in fluorinated oil with 5.0 mM PFPE(H)-Tris. However, if we swap out the surfactant for PFPE(H)$_2$-ED$_{900}$, no interfacial wrinkle was formed after the solvent extraction (**Figure 2 b**). In line with the observations in **Figure 1 d**, we also performed wrinkle tracking for droplets containing 1.0 wt% CdS@SiO$_2$ NPs in fluorinated oil with 5.0 mM PFPE(H)-Tris. As shown in **Figure 2 c**, no interfacial wrinkle was formed as expected.

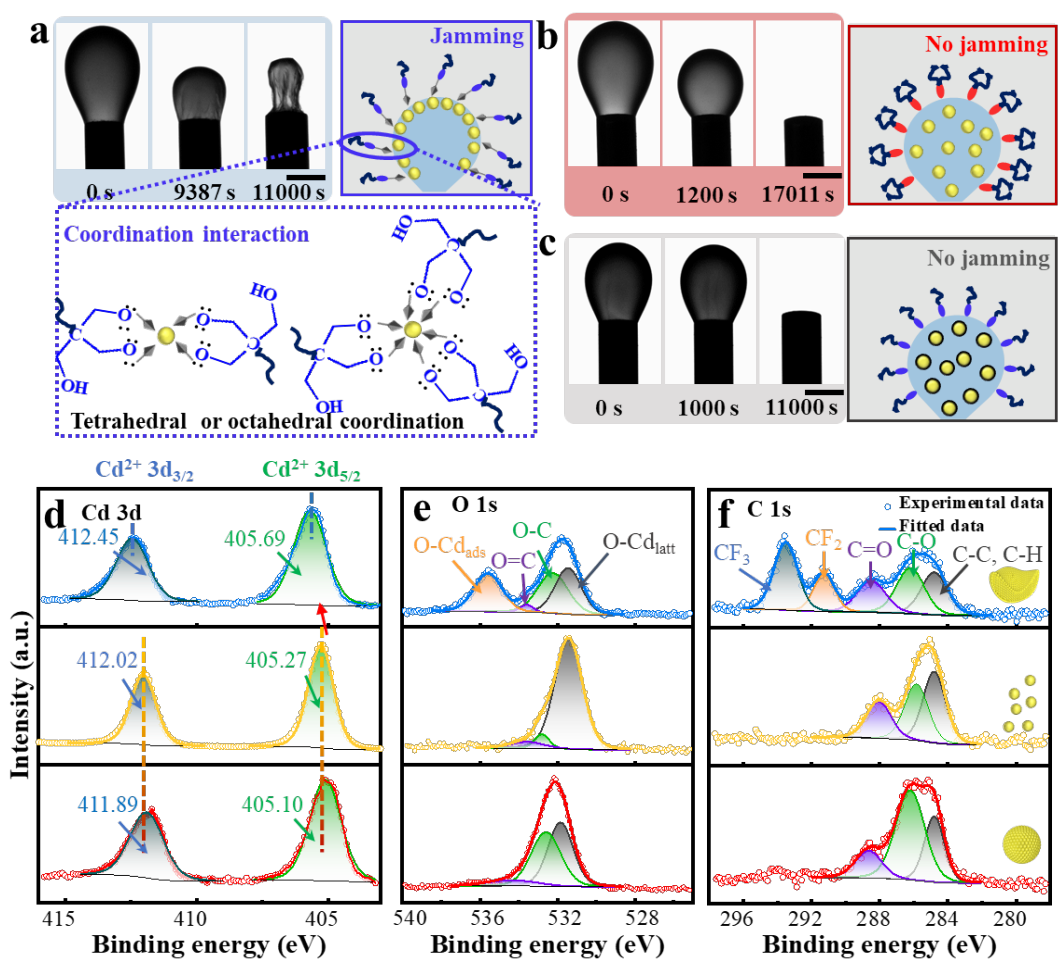

*Figure 2.* Mechanism of the formation, assembly and interfacial jamming of CdS NPs at the mixture solution-fluorinated oil interface. a-c) Morphology evolution of the pendant droplet with or without the interfacial jamming. The concentration of CdS NPs (shown in yellow) and CdS@SiO$_2$ NPs (shown as yellow circles with black stroke) is 1.0 wt% and the concentration of fluorosurfactants are 5.0 mM. Scale bars are 500 μm in (a-c). d-f) XPS spectra of CdS NPs, CdS NP colloidosomes and CdS NP supraparticles: d) Cd 3d; e) O 1s; f) C 1s. Note: Lattice Cd-O bond (O-Cd$_{latt}$) and adsorbed oxygen species (O-Cd$_{ads}$) on the surface represent the intrinsic Cd-O lattice bonding and the coordination interaction between fluorosurfactants and CdS NPs, respectively, as shown in panel (e).

In addition, wrinkling was observed at the interface of droplets containing 1.0 wt% ZnS NPs after solvent extraction, regardless of whether the fluorosurfactants used were PFPE(H)$_2$-ED$_{900}$ or PFPE(H)-Tris. In contrast, no wrinkle was observed at the droplet interface when the droplet contained the ZnS@SiO$_2$ NPs, regardless of the fluorosurfactant used PFPE(H)$_2$-ED$_{900}$ or PFPE(H)-Tris. These results were shown in **Figure S4 a-b**.

These results reveal that the hydroxyl group in PFPE(H)-Tris can effectively trap the CdS NPs at the droplet interface. Similarly, both the PFPE(H)-Tris and PFPE(H)$_2$-ED$_{900}$ can trap the ZnS NPs at the droplet interface, leading to the formation of CdS NP-CS or ZnS NP-CS with concave morphology. In the absence of interfacial jamming, structures such as CdS NP-SP, CdS@SiO$_2$ NP-SP and ZnS@SiO$_2$ NP-SP are constructed instead.

To further investigate the interactions causing the interfacial jamming, we measured the zeta potentials of these colloids, including CdS NPs, ZnS NPs, CdS@SiO$_2$ NPs and ZnS@SiO$_2$ NPs. All measured zeta potentials were negative, indicating that the particles carry a net negative surface charge, as shown in **Table S1**. The fluorosurfactants used in this study are nonionic. Therefore, electrostatic interactions between the nanoparticles and fluorosurfactants are unlikely to play a significant role in interfacial jamming. However, interactions between the hydroxyl groups and MS NPs are effectively suppressed by coating the NPs with a SiO$_2$ layer, as demonstrated in the **Figure 1 d**. Consequently, we hypothesize that the interfacial jamming may be induced by coordination interactions between the fluorosurfactants and MS NPs.

XPS is widely used to study coordination interactions between metal ions and organic components by detecting shifts in binding energies, which are indicative of such interactions[47]. Accordingly, we characterized CdS NPs, CdS NP-SPs, and CdS NP-CSs using XPS, focusing on the Cd 3d, O 1s, and C 1s spectra (**Figure 2 d-f**). In the XPS spectra, the red, yellow, and blue lines represent CdS NP-SPs, CdS NPs, and CdS NP-CSs, respectively. Compared to the CdS NPs, the binding energy of Cd$^{2+}$ 3d$_{3/2}$ and Cd$^{2+}$ 3d$_{5/2}$ in CdS NP-CSs and CdS NP-SPs exhibited slight shifts (**Figure 2 d**). These shifts indicate the electron transfer from the fluorosurfactant to Cd$^{2+}$, resulting in the positive shift due to the high electronegativity of the hydroxyl group[47, 48]. **Figure 2 e** demonstrates an emerging peak corresponding to O-Cd$_{ads}$ in the XPS spectra of CdS NP-CSs (blue line), in comparison to CdS NPs (yellow line) and CdS NP-SPs (red line), which only show the O-Cd$_{latt}$ peak. This further indicates the connection of the hydroxyl group to the CdS NPs through coordinate interactions in a form of the Cd-O coordinated bond[49]. Importantly, the peak is not observed in the XPS spectra of CdS NP-SPs, suggesting there is no or minor interaction with between the NPs and fluorosurfactants. Finally, the peaks corresponding to $CF_3$ and $CF_2$ in the C 1s XPS spectra are observed in CdS NP-CSs, but not in CdS NP-SPs (**Figure 2 f**), suggesting the fluorosurfactant was adsorbed at the interface of CdS NP-CSs. The results are also consistent with our previous work on the $Cu_xO$ NP-SPs and $Cu_xO$ CSs formed using similar fluorosurfactants[37]. These findings suggest that the hydroxyl group from PFPE(H)-Tris is more active in forming coordinate interaction compared to the amide bond from PFPE(H)$_2$-ED$_{900}$ with CdS NPs, resulting in the distinct construction of CdS NP-CSs and CdS NP-SPs, respectively. However, the coordination number of $Cd^{2+}$ with fluorosurfactant PFPE(H)-Tris remains to be investigated. Possible coordination geometries, including tetrahedral and octahedral configurations, are proposed in Figure **Figure 2 a**.

In addition, XPS was employed to investigate the coordination interactions between fluorosurfactants and ZnS NPs, as shown in **Figure S4 c-d**. The result regarding interfacial jamming aligned with our previous discussion (**Figure S3**), confirming our hypothesis of coordination interactions. The binding energy shifts were consistent with those observed for CdS NPs with fluorosurfactants.

Based on our investigation, we demonstrate that coordination interactions between the metal ion in the MS NPs are an indispensable factor in the evaporation-induced confined self-assembly. However, to the best of our knowledge, these interactions have not been discussed in the literature, particularly in studies focused on supraparticle formation. We suggest that although interfacial interactions may exist, high initial concentrations of colloids could obscure the effects of surface consolidation during drying due to particle crowding. In addition, the self-assembly behavior of CdS NPs and ZnS NPs exhibited differences in coordination strength with the fluorosurfactants, resulting in distinct morphologies.

**Relationship of assembly morphology with interfacial interaction and nanoparticle concentration**

We designed a set of experiments to examine how colloids crowding affects the final shape of colloidal self-assemblies. Microdroplets containing CdS NPs were generated and deposited onto a Petri dish using two different fluorosurfactants, and the self-assembly process was monitored *via* optical microscopy. Representative time-lapse snapshots are shown in **Figure 3 a**. The size of these CdS NP-SPs trends upward with the increased concentration of CdS NPs, while the shrinkage ratio from the microdroplet to SPs decreases, as depicted in **Figure 3 b**. The results are consistent with previous reports on evaporation-induced confined self-assembly and the correlation between colloids concentration and final structure[26, 50, 51]. In contrast, considering the coordination interaction between the CdS NPs and fluorosurfactants PFPE(H)-Tris, we expect the self-assembly to occur under condition of the CdS NPs jamming at the droplet interface, subsequently altering the morphology of the assembly. However, with increasing initial CdS NP concentrations, more and more self-assemblies do not exhibit a complete collapse of the outer colloidosome (**Figure 3 c)**, showcasing diversity of the morphology. We further counted the morphologies of observed assemblies, as shown in **Figure 3 d** with a histogram. It showed that at the concentrations of colloids above 10% the formation of CdS NP-SPs fully dominates the population of the final self-assembly completely, thereby concealing the presence of the surface effects. However, there is a difference between the optical microscopic images and SEM images (Figure 3 c, t = final *vs* SEM) of final assemblies. We attribute it to the influence of evaporation speed on the morphology under the interfacial interactions, which is a kinetically controlled process. This influence is distinct from the impact of evaporation speed on the internal crystalline lattice structure of supraparticles. While evaporation speed also affects their internal configuration of supraparticles, their overall shape remains spherical[30].

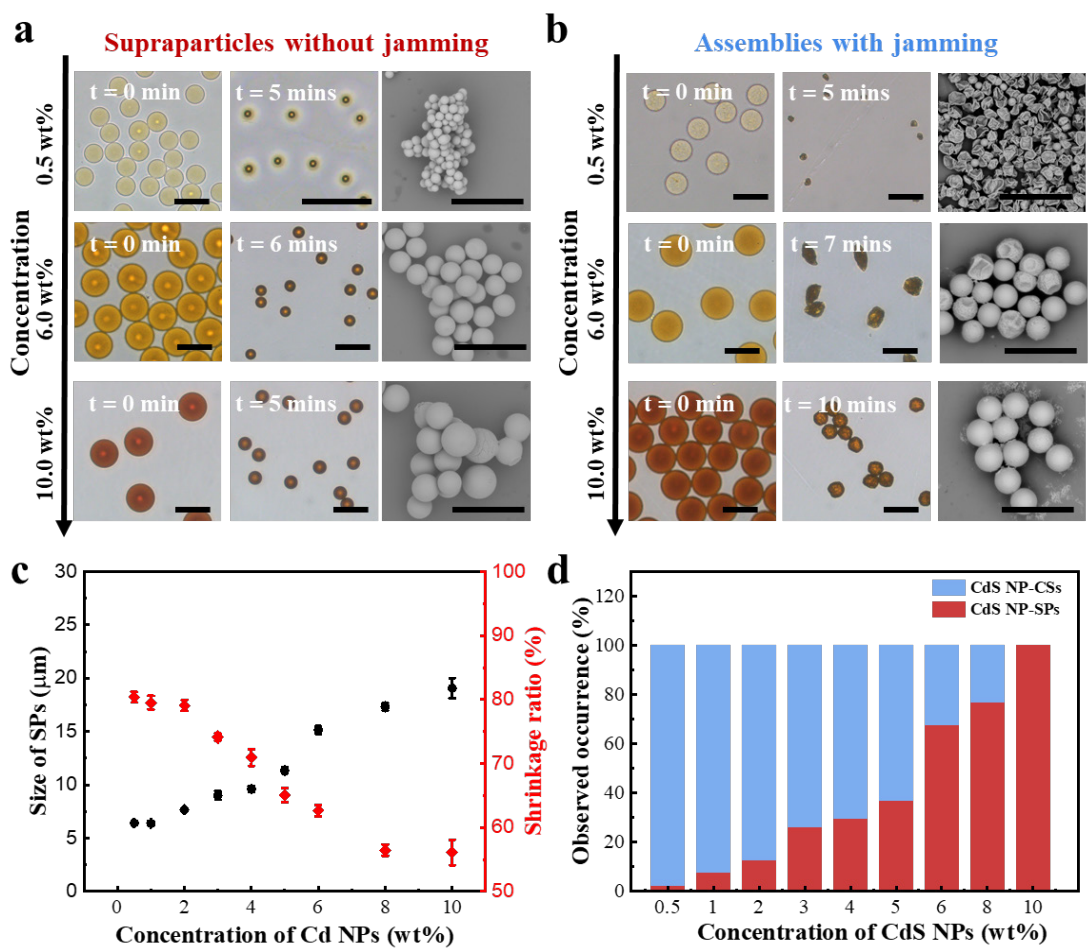

*Figure 3.* *Effects of the interfacial jamming and concentrations of CdS NP on the morphologies of CdS NP self-assemblies. a) OM and SEM images showing the formation of CdS NP-SPs at various concentrations of CdS NP, stabilized by 5.0 mM PFPE(H)$_2$-ED$_{900}$. b) OM images and SEM images of CdS NP-CSs and CdS NP-SPs at different concentration of CdS NP, stabilized by 5.0 mM PFPE(H)-Tris. (c) The average size and shrinkage ratio of CdS NP-SPs formation as increasing CdS NP concentrations, corresponding to OM and SEM images of (a). d) Statistical evaluation of the ratio of CdS NP-CSs to CdS NP-SPs observed with increasing the concentration of CdS NPs, corresponding to SEM images of (c). Note: For Figure 3b, we analyzed over 10 clusters and 10 microdroplets to obtain average size plots and shrinkage ratios; for Figure 3d, we examined 150 clusters to obtain statistical evaluation. Scale bars are 50 μm in the images of (a) and (c).*

The phenomenon in **Figure 3** was also observed for ZnS NPs self-assembling in the presence of PFPE(H)$_2$-ED$_{900}$ (**Figure S5 a**). The morphologies transitioned from the ZnS NP-CSs at low initial concentration of colloids to ZnS NP-SPs at higher concentration (5%). The corresponding occurrence of the observed morphologies were calculated in **Figure S5 b**. It demonstrates that the weak coordination interaction between the ZnS NPs and PFPE(H)$_2$-ED$_{900}$ results in similar regularity when compared to the CdS NPs and PFPE(H)-Tris combination. We here hypothesize that the coordination interactions act as an adsorption with monolayer and be limited to the droplet interface. The behavior of ZnS nanoparticles self-assembling into concave colloidosomes at low concentrations and supraparticles at high concentrations of ZnS is consistent with that observed with the commercial fluorosurfactant (Pico surf $^{TM}$) (**Figure S6 a-b**). We suspect that similar effect of coordination interactions is governed in the formation of these self-assemblies, although the exact composition of the Pico surf $^{TM}$ is not known. Similar to the CdS nanoparticles, the coordination interaction was eliminated by coating the ZnS particles with a SiO$_2$ layer (**Figure S6 c**). This similarity could help explain why Zn-based metal-organic framework particles are trapped at the interface[6, 35]. The coordination interaction between the ZnS NPs and PFPE(H)-Tris is expected to be stronger than PFPE(H)$_2$-ED$_{900}$, which was confirmed by the assemblies exhibited concave colloidosomes and the thickness increases, as shown in **Figure S7 a-i**. It should be noted that the ZnS NPs dispersed particles into the fluorinated oil, owing to the PFPE(H)-Tris surface-modified particles to leave the aqueous of droplet. This suggests that the coordination interaction with PFPE(H)-Tris is so strong that it allows them to cross the boundary of the droplet and get dispersed into the oil phase.

**Structuring binary colloid assembly with controllable segregation**

We noticed that the morphologies of assemblies can be tuned by the interfacial interactions between the MS NPs and fluorosurfactants, and concentrations of nanoparticles in suspension. Based on these findings, we hypothesize that using binary colloids with different affinities to the fluorosurfactant within a microdroplet can lead to a novel way of engineering the internal structure of the colloidal self-assembly[32, 33]. In addition, the co-self-assembly of binary nanoparticles within microdroplets has attracted interest for its applications in porous materials and colloidal segregation[24, 34].

We conducted an experiment to self-assemble binary nanoparticles to attempt engineering core-shell structures based on the affinity of the colloids to the fluorosurfactant. In this experiment, we used CdS NPs and SiO$_2$ NPs coated with carbosilane (SiO$_2$-C$_8$ NPs), using droplet templating techniques stabilized by different fluorosurfactants. The carbosilane coating on the SiO$_2$ NPs eliminates hydroxyl groups, to prevent coordination interactions between the CdS NPs and SiO$_2$ NPs. This causes the nanoparticles to stay in suspension during the self-assembly. The coordination interaction between CdS NPs and hydroxyl group of the fluorosurfactant PFPE(H)-Tris ensured a core/shell binary NPs assembly, as depicted in the schematic illustration (**Figure 4 a**). Due to the low concentration of NPs during the assemble process (**Video S4**), the assembly still exhibited a concave morphology, as shown in **Figure 4 b-c**. However, the SEM image and the FIB-SEM cross-section (**Figure 4 d-e**) still demonstrates selective absorption CdS NPs at the droplet interface, which can be seen as a single layer of CdS NPs on the outer surface of the assembly. Interestingly there are still CdS particle inclusions in the volume of the SiO$_2$-C$_8$ NPs (**Figure 4 e**). We believe these inclusions can be reduced by tuning down the initial concentration of the CdS NPs. As a control we prepared a binary nanoparticle assembly stabilized by PFPE(H)$_2$-ED$_{900}$, which was supposed to not have any interactions with the colloids. The self-assembly exhibited a spherical morphology with homogeneous distribution of both CdS NPs and SiO$_2$-C$_8$ NPs, as depicted in the schematic illustration of without jamming in **Figure 4 a.** The time-lapse of optical microscopic images were shown in **Figure 4 f-g** and further demonstrated in **Video S5**. This self-assembly as expected was formed without the interfacial jamming leading to a spherical isotropic assembly. The top view and cross-sectional of SEM images, shown in **Figure 4 h** and **Figure 4 i**, illustrate the homogeneous distribution of the CdS NPs and SiO$_2$-C$_8$ NPs both at the surface and the interior of the assembly, consistent with our findings.

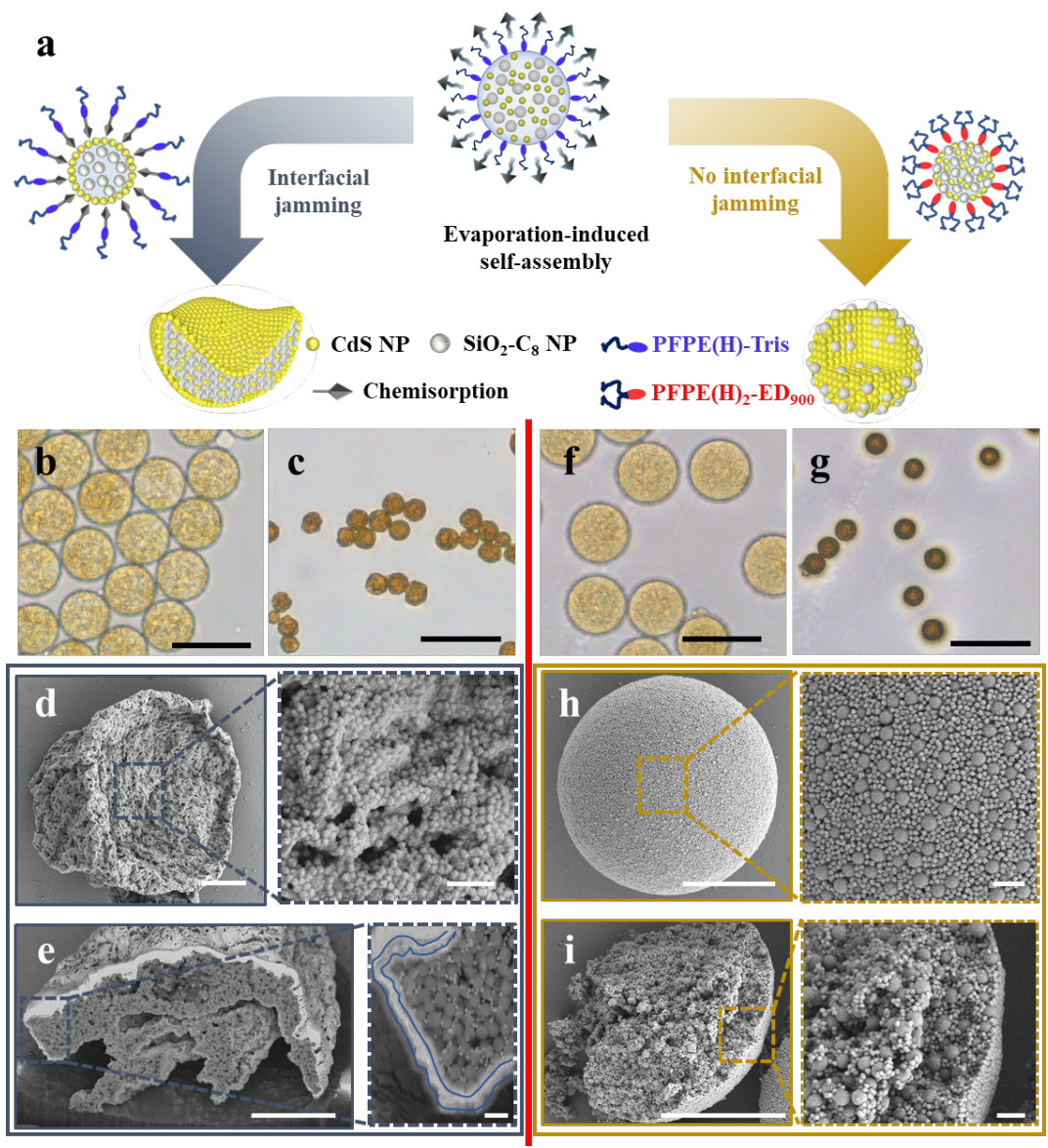

***Figure 4.*** *Binary colloid self-assemblies with controllable segregation. a) Schematic representation of binary nanoparticles self-assembly, highlighting selective coordination interactions between CdS NPs or SiO$_2$-C$_8$ NPs and the stabilizing fluorosurfactant. The left panel depicts the co-self-assembly of binary colloids (CdS NPs and SiO$_2$-C$_8$ NPs) within a microdroplet, stabilized by the fluorosurfactant PFPE(H)-Tris; the right panel depicts the co-self-assembly of binary colloids (CdS NPs and SiO$_2$-C$_8$ NPs) within a microdroplet, stabilized by the fluorosurfactant PFPE(H)$_2$-ED$_{900}$. b-c) Optical images of the start and the end of the binary colloids self-assembly in microdroplets, stabilized by PFPE(H)-Tris. The concentration of the SiO$_2$-C$_8$ NPs was insufficient to form a dense spherical assembly. d) The SEM of the top view of a binary assembly from panel c) with a zoomed-in view of the surface shown in the right panel. e) The side view of a FIB cut through a binary assembly with a detailed view of interface and interior shown in the right panel. f-g) Optical images of the start and end of binary colloids self-assembly in microdroplets, stabilized by PFPE(H)$_2$-ED$_{900}$. h) The SEM image of binary*
17

*assembly with a zoomed-in view of the surface shown in the right panel. i) The side view of a ruptured binary assembly with a detailed view of interface and interior shown in the right panel. Scale bars: 50 μm in these images of b-c and f-g, 5 μm in main panel of d-e and i-j, and 500 nm in the zoomed-in panel of d-e and i-j.*

In addition, we have performed a similar assembly based on ZnS NPs and SiO$_2$-C$_8$ NPs, which as well showed the expected phenomenon with ZnS NPs segregating at the interface, shown in **Figure S8**. These results highlight the role of coordination interaction between metal-based colloids and fluorosurfactants in shaping a colloidal self-assembly.

## Conclusion

We studied the effect of interfacial interactions on the morphologies of colloidal assemblies, facilitated by evaporation-induced droplet templating. Our study confirms that interfacial interactions between fluorosurfactants and metal-based colloids, and concentrations of metal-based colloids are important factors determining the morphology and the structure of the final assembly. Our research suggests that the coordination interaction between the -OH groups of the surfactant and metal ions in the colloid is the reason of the adsorption of the colloids to the surface of the droplet during the self-assembly. Experiments with assemblies of microdroplets containing CdS NPs and ZnS NPs and also core-shell nanoparticles CdS@SiO$_2$ and ZnS@SiO$_2$ stabilized by fluorosurfactants PFPE(H)$_2$-ED$_{900}$ or PFPE(H)-Tris, offered detailed insights into these surface interactions. The SiO$_2$ coating on the nanoparticles allowed to mitigate the coordination interaction between the fluorosurfactant and MS NPs. In addition, the pendant drop technique showed that the MS NPs were jammed at the droplet interface, whereas MS@SiO$_2$ NPs were not. Further evidence for the nature of these interactions was provided by measurements of the zeta potentials of the colloids and using XPS on the final assemblies. These measurements illustrated that interfacial jamming is facilitated by coordination interactions and not the charge-based interactions. Finally, we used the discovered surface interactions to engineer different morphologies of binary colloidal assemblies, featuring diverse distribution of colloids (core/shell or homogenous), have been structured *via* self-assembling the MS NPs and SiO$_2$-C$_8$ NPs within microdroplets, assisted by different fluorosurfactants. This study provides insights into the chemical dynamics at play during droplet templating processes and underscores the importance of considering both physical and chemical interactions in the design of advanced colloidal assembled materials.

**Materials and Method**

**Materials and reagents**

All chemicals were used as provided without further purification unless noted otherwise. Zinc nitrate hexahydrate (Zn(NO$_3$)$_2$·6H$_2$O, ≥ 99%), cadmium nitrate tetrahydrate (Cd(NO$_3$)$_2$·4H$_2$O, ≥ 99%), tetraethyl orthosilicate (TEOS, ≥ 99.99%), and ammonia solution (25%-28%) were obtained from Aladdin (China). Diethylene glycol (DEG, ≥ 99%) was purchased from Energy chemical (China). DuPont Krytox 157-FSH (carboxylated perfluoropolyether, PFPE(H)-COOH, $M_w$=7000~7500 g·mol$^{-1}$) was purchased from Miller-Stephenson Chemical Co. Inc. (Danbury, CT, USA). 3M Novec$^{TM}$ HFE 7100 and 3M Novec$^{TM}$ HFE 7500 were purchased from Miller- 3M (St. Paul, MN, USA). Jeffamine ED$_{900}$, triethylamine (TEA, ≥99.5%), N-methylmorpholine (NMM, 99%), benzotrifluoride (BTF), Tris(hydroxymethyl)aminomethane (Tris, ≥ 99%), thiourea (TU) and 3-glycidyloxypropyltrimethoxysilane (≥ 97%) were obtained from Sigma-Aldrich (Shanghai, China).

**Synthesis of fluorosurfactants**

The fluorosurfactants, including PFPE(H)$_2$-ED$_{900}$ and PFPE(H)-Tris, were synthesized following the methodology outlined in our previous work[37]. In brief, the synthesis involved mixing PFPE(H)-COOH (1.00 mmol) and N-methylmorpholine (NMM, 1.20 mmol) in 3M Novec HFE 7100 (10.0 mL). Ethyl chloroformate (1.20 mmol) was then gradually added to this mixture at 0 ºC, and the reaction was allowed to proceed for 0.5 h. Subsequently, this mixture was introduced into another solution containing Jeffamine ED$_{900}$ or Tris(hydroxymethyl)aminomethane (Tris, 0.50 mmol) and triethylamine (1.20 mmol) in benzotrifluoride (BTF, 6.0 mL). The combined mixtures were stirred for 10 minutes at 0 ºC, followed by a reaction period of 0.5 h at room temperature (RT) with continuous stirring. The synthesized fluorosurfactants were purified by washing with ethanol five times, and dried under vacuum at 60 ºC for 24 h to yield a transparent, viscous liquid.

**Synthesis of metal-based colloids and metal@SiO$_2$ colloids**

For CdS NPs synthesis, we used the previously reported protocol[52]. Firstly, the seed of CdS NPs was prepared by hot-injecting 100 µL of the stock solutions (1.0 M Cd(NO$_3$)$_2$·4H$_2$O and 1.0 M TU) into a three-necked flask containing 30 mL of DEG solution and 0.6 g of PVP. The mixture was heated to 180 ºC with stirring and maintained for 5 min. The flask was then rapidly cooled in an ice-water bath while stirring continued. For subsequent seed growth, 0.616 g Cd(NO$_3$)$_2$·4H$_2$O and 0.152 g TU powder were added to the cooled solution. The mixture was heated to 160 ºC over 16 min and held at this temperature for 1 h. After the reaction, the CdS NPs allowed to cool down to RT and purified by centrifugation and washed three times with deionized water and ethanol, respectively.

For the synthesis of ZnS NPs, we used the previously reported protocol[53]. Firstly, the seed of ZnS NPs was prepared through hot-injecting 400 µL of a 2.0 M TU stock solution into a three-necked flask containing 0.119 g Zn(NO$_3$)$_2$·6H$_2$O, 30 mL of DEG solution and 0.6 g of PVP. The mixture was heated to 160 ºC with stirring for 10 min. Subsequently, the flask was quickly cooled in an ice-water bath while maintaining stirring. For subsequent seed growth, 3.570 g of Zn(NO$_3$)$_2$·6H$_2$O, 1.83 g of TU powder and 1.2 g of PVP were dissolved in 60 mL of DEG at 70 ºC under vigorous magnetic stirring to form a transparent solution. Then, 2 mL of the aforementioned seed solution was injected into the reaction solution, which was further heated to 160 ºC and maintained at this temperature for 2 h under magnetic stirring. The ZnS NPs were purified by centrifugation and washed three times with deionized water and ethanol after cooling to RT.

For metal@SiO$_2$ colloids synthesis, we followed the reported protocol outlined in reference[54]. Initially, 0.1 g of CdS NPs or 0.4 g of ZnS NPs were dispersed in 40 mL of ethanol. Subsequently, 5.0 mL DI water and 0.75 mL ammonia solution were added and the mixture was further stirred. Then, 0.6 mL TEOS was injected quickly, and stirred for 2 h at RT. The metal@SiO$_2$ colloids were purified by centrifugation and washed five times with deionized water.

For the synthesis of silica nanoparticles coated with carbosilane (SiO$_2$-C$_8$ NPs), SiO$_2$ NPs were first produced using a modified Stöber method[55]. Initially, the solution A was prepared by stirring together 65 mL of absolute ethanol, 15 mL of ammonia solution and 100 mL of DI water. Concurrently, the solution B was prepared by mixing 185 mL of absolute ethanol with 10 mL of TEOS. Solutions A and B were then combined and stirred at 300 rpm for 1.0 h at RT. The resulting mixture was centrifuged and washed three times sequentially with ethanol and water, followed by drying. Subsequently, 3-glycidyloxypropyltrimethoxysilane was covalently bonded onto the SiO$_2$ NPs surface through hydroxyl group[56]. The above-synthesized SiO$_2$ NPs dispersed into 100 mL of toluene and sonicated for 0.5 h; then, 4 mL 3-glycidyloxypropyltrimethoxysilane was added to the SiO$_2$ solution and refluxed 24 h at 110 ºC under the N$_2$ atmosphere. The SiO$_2$-C$_8$ NPs collected through centrifugation and washed three times with ethanol.

**PDMS droplet generator fabrication**

The polydimethylsiloxane (PDMS) droplet generator with a flow-focusing geometry was fabricated according to methods described in our previous work[37]. Using standard soft lithography techniques, a mold was created from a silicon wafer coated with SU-3050 photoresist to design the channel structure. The PDMS pre-polymer was then poured onto the silicon mold in a 10:1 mass ratio of base to curing agent. This mixture was stirred and degassed in a vacuum chamber to remove any internal bubbles. Both the PDMS slide and the glass cover slide were treated in a plasma cleaner (PDC-002, Mycro Technologies Co., Ltd., Beijing, China) at 10 W for 60 seconds and subsequently bonded face-to-face to form the microfluidic chip. Finally, to achieve a hydrophobic microchannel, Aquapel (PPG Industries, PA, USA) was injected, followed by air flushing and heating at 120 ºC for 0.5 h.

**Construction of assemblies *via* droplet microfluidics.**

Microdroplets containing colloids were generated using a flow-focus junction chip with two inlets and one outlet. The inlet, connected to a syringe pump, was used for the aqueous metal-based colloid-laden phase (dispersed phase), while the other inlet was used for injecting the fluorinated phase, comprising either 5.0 mM PFPE(H)$_2$-ED$_{900}$ or 5.0 mM PFPE(H)-Tris (continuous phase). The flow rate of two syringes is set as 5.0 µL·min$^{-1}$ for continuous phases and 2.5 µL·min$^{-1}$ for dispersed phase and the size of microdroplets can be adjusted by independently altering the flow rate. The generation of 1:1 water-ethanol (v/v) mixture-in-fluorinated oil (M/O$_F$) droplet containing MS NPs at a concentration of 1.0 wt/v% were observed and recorded using an inverted optical microscope (Olympus IX2, Tokyo, Japan) equipped with a high-speed camera (Phantom MIRO M110, Vision Research Inc., Wayne County, NC, USA) in bright-field mode. After microdroplet formation, the M/O$_F$ microdroplets were subsequently collected for 30 mins and stored the microdroplets in glass vials at 33 ºC under ambient atmosphere conditions. Meanwhile, the microdroplets were placed under an inverted optical microscope (Olympus IX2, Tokyo, Japan) to observe droplet shrinkage during solvent evaporation. The microdroplets were allowed to dry over approximately five days. Subsequently, the obtained assemblies were purified by washing with Novec$^{TM}$ HFE 7100 oil for four times, and were stored in Novec$^{TM}$ HFE 7100 oil for further characterization.

**Analytic methods**

**Fourier transform infrared.** FT-IR spectra were recorded on a Vertex 70 FT-IR spectrometer (Bruker, Germany) in the transmittance mode. The wavenumber range was 400-4000 cm$^{-1}$, and the resolution was 2 cm$^{-1}$.

**Interfacial rheology measurements.** Interfacial rheology behavior was performed using a pendant drop technique using OCA 15 Pro instrument (Dataphysics, Germany) at ambient temperature. Colloid NPs were dispersed in a 1:1 water-ethanol (v/v) mixture with a concentration of 1.0 wt/v% MS NPs to form the dispersed phase, while fluorinated oil (Novec$^{TM}$ HFE 7500) containing the fluorosurfactant at the concentration of 5.0 mM (above the critical micelle concentration (CMC)) served as the continuous phase. The dispersed phase was injected slowly into the continuous phase using a syringe pump to maximize droplet volume. The drop was captured on a camera to record changes of interfacial rheology.

**Zeta potential measurements.** The zeta potential measurements were performed using a NanoBrook 90Plus PALS instrument (Brookhaven, USA). All colloids were dispersed in ethanol, and each measurement was repeated three times to minimize instrumental error.

**X-ray photoelectron spectroscopy measurements.** The composition of the synthesized assemblies was analyzed using XPS equipment (ESCALAB 250, USA). In-depth XPS profiles were obtained by argon-ion beam etching at a rate of 0.06 nm·s$^{-1}$. All XPS spectra were calibrated to the carbon peak (C 1s, 284.8 eV). The XPS peaks were fitted using CasaXPS software, ensuring that the full widths at the half-maximum (FWHM) of the main peaks were kept below 3.0 eV, with a fixed Lorentzian/Gaussian ratio of 20%.

**Scanning electron microscopy and focused-ion beam scanning electron microscopy sample preparation and measurement.** To prepare the sample for Scanning electron microscopy (SEM), 10 μL of CdS NPs, ZnS NPs, CdS@SiO$_2$ NPs, ZnS@SiO$_2$ NPs, SiO$_2$-C$_8$ NPs or assemblies (including single colloids and bi-colloids assemblies) were drop-casted on a silicon wafer, and subsequently were dried under ambient condition. The structure of assemblies was characterized by GeminiSEM 500 (Carl Zeiss Microscopy GmbH) with an accelerating voltage of 2.0 kV in secondary electron imaging mode. To further analyze the interior morphology of bi-colloids assemblies, bi-colloids assemblies were sectioned by focused ion beam (FIB, Thermo Scientific Helios 5 UX) using a Ga ion source. The milled cross-section was imaged in secondary electron mode at an accelerating voltage of 5 kV.

**Acknowledgements**

We acknowledge financial support from the Key Project of the National Natural Science Foundation of China (No. 12131010), the International Cooperation Base of Infrared Reflection Liquid Crystal Polymers and Device (2015B050501010), the Science and Technology Project of Guangdong Province (No. 2024A1515010687, 2023A1515010935), and the China Scholarship Council (CSC) (Grant No. 202106750031).

**Contributions**

J. Yao, L. Shui, and S. Pud conceived and designed the experiments and worked on the manuscript. J. Yao, S. Xie, S. Huang, and Z. Liu conducted the droplet microfluidics experiments. J. Yao, S. Xie and W. Xu performed the colloidal self-assembly experiments, and W. Xu draw the schematic figures. J. Yao, S. Xie, M. Jin, and L. I Segerink performed data analysis. J. Yao, S. Xie, S. Huang, M. Jin, L. I. Segerink, and S. Pud, L. Shui wrote and revised the manuscript.

# Reference:

(1) Posnjak, G.; Yin, X.; Butler, Paul.; Bienek, O,; Dass, M,; Lee, S,; Liedl, T. Diamond-lattice photonic crystals assembled from DNA origami. *Science* **2024**, *384* (6697), 781–785.
(2) Williams, C. A.; Parker, R. M.; Kyriacou, A.; Murace, M.; Vignolini, S. Inkjet printed photonic cellulose nanocrystal patterns. *Advanced Materials* **2024**, *36* (1), e2307563.
(3) Li, Z,; Fan, Q,; Yin, Y. Colloidal self-assembly approaches to smart nanostructured materials. *Chemical reviews* **2021**, *122* (5), 4976-5067.
(4) Bian, F.; Sun, L.; Cai, L.; Wang, Y.; Wang, Y.; Zhao, Y. Colloidal crystals from microfluidics. *Small* **2019**, *16* (9), e1903931.
(5) Xu, Z.; Hueckel, T.; Irvine, W. T.; Sacanna, S. Caged colloids. Chemistry of Materials 2023, 35 (16), 6357-6363.
(6) Fujiwara, A.; Wang, J.; Hiraide, S.; Gotz, A.; Miyahara, M. T.; Hartmann, M.; Apeleo Zubiri, B.; Spiecker, E.; Vogel, N.; Watanabe, S. Fast gas-adsorption kinetics in supraparticle-based MOF packings with hierarchical porosity. *Advanced Materials* **2023**, *35* (44), e2305980.
(7) Yao, J.; Liu, Z.; Jin, M.; Zou, Y.; Chen, J.; Xie, P.; Wang, X.; Akinoglu, E. M.; Zhou, G.; Shui, L. Uniform honeycomb CNT-microparticles prepared via droplet-microfluidics and sacrificial nanoparticles for electrochemical determination of methyl parathion. *Sensors and Actuators B: Chemical* **2020**, *321*, 128517.
(8) Wang, J.; Le-The, H.; Wang, Z.; Li, H.; Jin, M.; van den Berg, A.; Zhou, G.; Segerink, L. I.; Shui, L.; Eijkel, J. C. T. Microfluidics assisted fabrication of three-tier hierarchical microparticles for constructing bioinspired surfaces. *ACS Nano* **2019**, *13* (3), 3638-3648.
(9) Zhu, P.; Wang, L. Microfluidics-enabled soft manufacture of materials with tailorable wettability. *Chemical Reviews* **2022**, *122* (7), 7010-7060.
(10) Wang, J.; Pinkse, P. W.; Segerink, L. I.; Eijkel, J. C. T. Bottom-up assembled photonic crystals for structure-enabled label-free sensing. *ACS Nano* **2021**, *15* (6), 9299-9327.
(11) Wang, J.; Eijkel, J. C. T.; Jin, M.; Xie, S.; Yuan, D.; Zhou, G.; van den Berg, A.; Shui, L. Microfluidic fabrication of responsive hierarchical microscale particles from macroscale materials and nanoscale particles. *Sensors and Actuators B: Chemical* **2017**, *247*, 78-91.
(12) Tang, Y.; Gomez, L.; Lesage, A.; Marino, E.; Kodger, T. E.; Meijer, J. M.; Kolpakov, P.; Meng, J.; Zheng, K.; Gregorkiewicz, T.; Schall, P. Highly stable perovskite supercrystals via oil-in-oil templating. *Nano Letters* **2020**, *20* (8), 5997-6004.
(13) Montanarella, F.; Urbonas, D.; Chadwick, L.; Moerman, P. G.; Baesjou, P. J.; Mahrt, R. F.; van Blaaderen, A.; Stoferle, T.;Vanmaekelbergh, D. Lasing supraparticles self-Assembled from nanocrystals. *ACS Nano* **2018**, *12* (12), 12788-12794.
(14) Xu, W.; Li, Z.; Yin, Y. Colloidal assembly approaches to micro/nanostructures of complex morphologies. *Small* **2018**, *14* (35), e1801083.
(15) Wang, J.; Le-The, H.; Shui, L.; Bomer, J. G.; Jin, M.; Zhou, G.; Mulvaney, P.; Pinkse, P. W. H.; van den Berg, A.; Segerink, L. I.; Eijkel, J. C. T. Multilevel spherical photonic crystals with controllable structures and structure-enhanced functionalities. *Advanced Optical Materials* **2020**, 8 (10), 1902164.
(16) Gu, X.; Liu, Y.; Chen, G.; Wang, H.; Shao, C.; Chen, Z.; Lu, P.; Zhao, Y. Mesoporous colloidal photonic crystal particles for intelligent drug delivery. *ACS Applied Materials & Interfaces* **2018**, *10* (40), 33936-33944.

(17) Wang, L.; Sun, L.; Bian, F.; Wang, Y.; Zhao, Y. Self-bonded hydrogel inverse opal particles as sprayed flexible patch for wound healing. *ACS Nano* **2022**, *16* (2), 2640-2650.

(18) Kim, Y. G.; Park, S.; Kim, S.-H. Designing photonic microparticles with droplet microfluidics. *Chemical Communications* **2022**, *58* (74), 10303-10328.

(19) Liu, D.; Aleisa, R.; Cai, Z.; Li, Y.; Yin, Y. Self-assembly of superstructures at all scales. *Matter* **2021**, *4* (3), 927-941.

(20) Li, W.; Zhang, C.; Wang, Y. Evaporative self-assembly in colloidal droplets: Emergence of ordered structures from complex fluids. *Advances in Colloid and Interface Science* **2024**, *333*, 103286.

(21) Wang, X.; Xu, Y.; Xiang, S.; Tao, S.; Liu, W. Hydrogel-assisted robust supraparticles evolved from droplet evaporation. *ACS Nano* **2024**, *18* (52), 35684-35695.

(22) Wang, J.; Hahn, S.; Amstad, E.; Vogel, N. Tailored double emulsions made simple. *Advanced Materials* **2022**, *34* (5), e2107338.

(23) Wang, J.; Sultan, U.; Goerlitzer, E. S. A.; Mbah, C. F.; Engel, M.; Vogel, N. Structural color of colloidal clusters as a tool to investigate structure and dynamics. *Advanced Functional Materials* **2019**, *30* (26), 1907730.

(24) Liu, W.; Midya, J.; Kappl, M.; Butt, H. J.; Nikoubashman, A. Segregation in drying binary colloidal droplets. *ACS Nano* **2019**, *13* (5), 4972-4979.

(25) Hagan, M. F.; Grason, G. M. Equilibrium mechanisms of self-limiting assembly. *Reviews of Modern Physics* **2021**, *93* (2), 025008.

(26) Wozniak, M.; Derkachov, G.; Kolwas, K.; Archer, J.; Wojciechowski, T.; Jakubczyk, D.; Kolwas, M. Formation of highly ordered spherical aggregates from drying microdroplets of colloidal suspension. *Langmuir* **2015**, *31* (28), 7860-7868.

(27) Jiang, X.; Jiang, B.; Mu, M.; Wang, T.; Sun, S.; Xu, J.; Wang, S.; Zhou, Y.; Zhang, J.; Li, W. Complex core-shell architectures through spatially organized nano-assemblies. *ACS Nano* **2025**, *19* (6), 6479-6487.

(28) Shah, R. K.; Kim, J. W.; Weitz, D. A. Monodisperse stimuli-responsive colloidosomes by self-assembly of microgels in droplets. *Langmuir* **2010**, *26* (3), 1561-1565.

(29) Liu, W.; Kappl, M.; Steffen, W.; Butt, H. J. Controlling supraparticle shape and structure by tuning colloidal interactions. *Journal of Colloid and Interface Science* **2022**, *607*, 1661-1670.

(30) Wang, J.; Mbah, C. F.; Przybilla, T.; Apeleo Zubiri, B.; Spiecker, E.; Engel, M.; Vogel, N. Magic number colloidal clusters as minimum free energy structures. *Nature Communicatios* **2018**, *9* (1), 5259.

(31) Thayyil Raju, L.; Koshkina, O.; Tan, H.; Riedinger, A.; Landfester, K.; Lohse, D.; Zhang, X. Particle size determines the shape of supraparticles in self-lubricating ternary droplets. *ACS Nano* **2021**, *15* (3), 4256-4267.

(32) Marino, E.; LaCour, R. A.; Moore, T. C.; van Dongen, S. W.; Keller, A. W.; An, D.; Yang, S.; Rosen, D. J.; Gouget, G.; Tsai, E. H. R.; Kagan, C. R.; Kodger, T. E.; Glotzer, S. C.; Murray, C. B. Crystallization of binary nanocrystal superlattices and the relevance of short-range attraction. *Nature Synthesis* **2023**, *3* (1), 111-122.

(33) Wang, D.; Dasgupta, T.; van der Wee, E. B.; Zanaga, D.; Altantzis, T.; Wu, Y.; Coli, G. M.; Murray, C. B.; Bals, S.; Dijkstra, M.; Blaaderen, van Blaaderen, A. Binary icosahedral clusters of hard spheres in spherical confinement. *Nature Physics* **2021**, *17* (1), 128–134.

(34) Liu, W.; Kappl, M.; Butt, H. J. Tuning the porosity of supraparticles. *ACS Nano* **2019**, *13* (12),

13949-13956.
(35) Wang, J.; Liu, Y.; Bleyer, G.; Goerlitzer, E. S. A.; Englisch, S.; Przybilla, T.; Mbah, C. F.; Engel, M.; Spiecker, E.; Imaz, I.; Maspoch, D.; Vogel, N. Coloration in supraparticles assembled from polyhedral metal-organic framework particles. *Angewandte Chemie International Edition* **2022**, *134* (16), e202117455.
(36) Roemling, L. J.; De Angelis, G.; Mauch, A.; Amstad, E.; Vogel, N. Control of buckling of colloidal supraparticles. arXiv 2024, 2409.00602.
(37) Yao, J.; Huang, S.; Xie, S.; Liu, Z.; Deng, Y.; Carnevale, L.; Jin, M.; Segerinkc, I. L.; Wang, D.; Shui, L.; Pud, S. A robust synthesis of fluorosurfactants with tunable functions via a two-step reaction. arXiv 2025, on hold.
(38) Tian, D.; Hao, R.; Zhang, X.; Shi, H.; Wang, Y.; Liang, L.; Liu, H.; Yang, H. Multi-compartmental MOF microreactors derived from pickering double emulsions for chemo-enzymatic cascade catalysis. *Nature Communications* **2023**, *14* (1), 3226.
(39) Chen, G.; Yu, Y.; Fu, X.; Wang, G.; Wang, Z.; Wu, X.; Ren, J.; Zhao, Y. Microfluidic encapsulated manganese organic frameworks as enzyme mimetics for inflammatory bowel disease treatment. *Journal of Colloid and Interface Science* **2022**, *607*, 1382-1390.
(40) Etienne, G.; Ong, I. L. H.; Amstad, E. Bioinspired viscoelastic capsules: Delivery vehicles and beyond. *Advanced Materials* **2019**, *31* (27), e1808233.
(41) De Angelis, G.; Gray, N.; Lutz‐Bueno, V.; Amstad, E. From surfactants to viscoelastic capsules. *Advanced Materials Interfaces* **2023**, *10* (13), 22024050.
(42) Feng, W.; Chai, Y.; Forth, J.; Ashby, P. D.; Russell, T. P.; Helms, B. A. Harnessing liquid-in-liquid printing and micropatterned substrates to fabricate 3-dimensional all-liquid fluidic devices. *Nature Communications* **2019**, *10* (1), 1095.
(43) Li, L.; Sun, H.; Li, M.; Yang, Y.; Russell, T. P.; Shi, S. Gated molecular diffusion at liquid-liquid interfaces. *Angewandte Chemie International Edition* **2021**, *60* (32), 17394-17397.
(44) Huo, Q.; Gao, Y.; Wu, W.; Hu, S.; Zhang, Z.; Li, Z.; Tian, Y.; Quan, P.; Li, W.; Liu, D. Colloidal jamming by interfacial self-assembled polymers: A robust route for ultrahigh efficient encapsulation. *Angewandte Chemie International Edition* **2022**, *61* (43), e202208738.
(45) Wang, B.; Yin, B.; Zhang, Z.; Yin, Y.; Yang, Y.; Wang, H.; Russell, T. P.; Shi, S. The assembly and jamming of nanoparticle surfactants at liquid-liquid interfaces. *Angewandte Chemie International Edition* **2022**, *61* (10), e202114936.
(46) Sun, S.; Li, S.; Feng, W.; Luo, J.; Russell, T. P.; Shi, S. Reconfigurable droplet networks. *Nature Communication* **2024**, *15* (1), 1058.
(47) Feltham, R. D.; Brant, P. XPS Studies of core binding energies in transition metal complexes. 2. Ligand group shifts. *Journal of the American Chemical Society* **1982**, *104*, 641-645.
(48) Ávila-Torres, Y.; Huerta, L.; Barba-Behrens, N. XPS-Characterization of heterometallic coordination compounds with optically active ligands. *Journal of Chemistry* **2013**, *2013* (1), 370637.
(49) Zhang, Y; Ji, Y.; Li, J.; Liu, H.; Hu, X.; Zhong, Z.; Su, F. Morphology-dependent catalytic properties of nanocupric oxides in the Rochow reaction. *Nano research* **2018**, *11*, 804–819.
(50) Yi, G. R.; Manoharan, V. N.; Klein, S.; Brzezinska, K. R.; Pine, D. J.; Lange, F. F.; Yang, S. M. Monodisperse micrometer-scale spherical assemblies of polymer particles. *Advanced Materials* **2002**, *14* (16), 1137-1140.
(51) Naveenkumar, P. M.; Roemling, L. J.; Sultan, U.; Vogel, N. Fabrication of spherical colloidal

supraparticles via membrane emulsification. *Langmuir* **2024**, *40* (42), 22245-22255.

(52) Su, X; Chang, J.; Wu, S.; Tang, B.; Zhang, S. Synthesis of highly uniform $Cu_2O$ spheres by a two-step approach and their assembly to form photonic crystals with a brilliant color. *Nanoscale* **2016**, *8* (11), 6155–6161.

(53) Fang, Q.; Wei, W.; Wu, C.; Su, X.; Wu, S. Synthesis of monodisperse colloidal ZnS spheres by an acid-free two-step approach. *European Journal of Inorganic Chemistry* **2023**, *26* (1), e202200486.

(54) Wei, W.; Wu, C.; Fang, Q.; Zhao, C; Su, X. Beyond cadmium yellow: CdS photonic crystal pigments with vivid structural colors. *Journal of Materials Chemistry C* **2023**, *11* (37), 12667-12674.

(55) Stöber, W.; Fink, A., Bohn, E. Controlled growth of monodisperse silica spheres in the micron size range. *Journal of Colloid and Interface Science* **1968**, *26* (1), , 62-69

(56) Ta, T. K. H.; Trinh, M-T.; Long, N. V.; Nguyen, T. T. M.; Nguyen, T. L. T.; Tran Linh Thuoc; Phan, B. T.; Mott, D.; Maenosono, S.; Tran-Van, H.; Le, V. H. Synthesis and surface functionalization of $Fe_3O_4$-$SiO_2$ core-shell nanoparticles with 3-glycidoxypropyltrimethoxysilane and 1,1′-carbonyldiimidazole for bio-applications. *Colloids and Surfaces A: Physicochemical and Engineering Aspects* **2016**, 504, 376-383.

# TOC

For Table of Contents Use Only

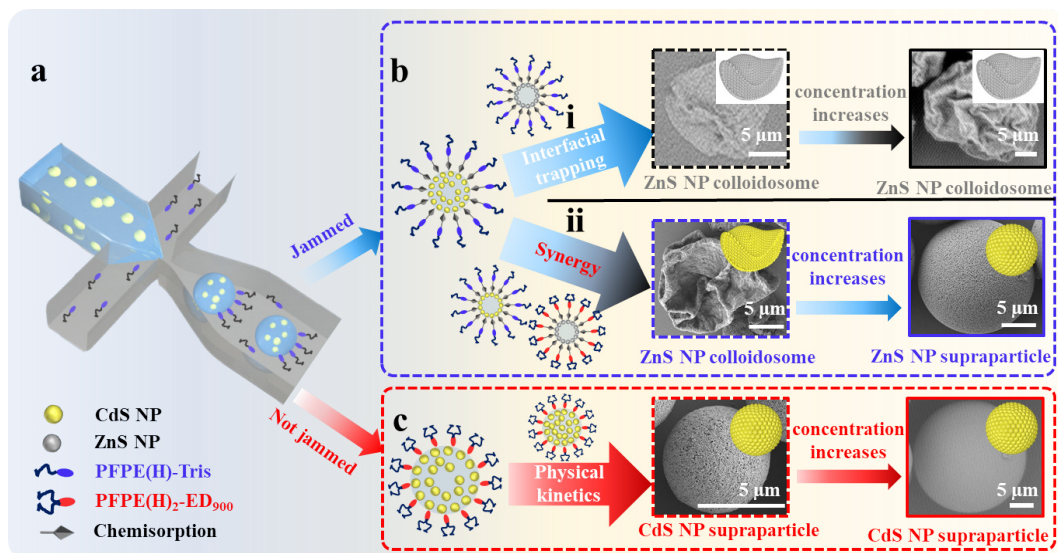

# SYNOPSIS

The interfacial coordination between the fluorosurfactant and metal-based colloids influences the morphology of metal-based colloidal self-assemblies to varying degrees during evaporation-driven self-assembly within microdroplets.